\documentstyle[epsfig,12pt,a4p]{article}

\newcommand{\kk}{\mbox{$K^+K^-$ }}
\newcommand{\ksks}{\mbox{$K^0_SK^0_S$ }}

\newcommand{\pipi}{\mbox{$\pi^+\pi^-$ }}

\begin{document}
\begin{titlepage}
\def\footnoterule{\hrule width 1.0\columnwidth}
\begin{tabbing}
put this on the right hand corner using tabbing so it looks
 and neat and in \= \kill
\> {10 February 1999}
\end{tabbing}
\bigskip
\bigskip
\begin{center}{\Large  {\bf A partial wave analysis of the centrally produced
$K^+ K^-$ and $K^0_SK^0_S$ systems
in $pp$ interactions at 450 GeV/c and new information on the spin of the
$f_J(1710)$}
}\end{center}
\bigskip
\bigskip
\begin{center}{        The WA102 Collaboration
}\end{center}\bigskip
\begin{center}{
D.\thinspace Barberis$^{  4}$,
W.\thinspace Beusch$^{   4}$,
F.G.\thinspace Binon$^{   6}$,
A.M.\thinspace Blick$^{   5}$,
F.E.\thinspace Close$^{  3,4}$,
K.M.\thinspace Danielsen$^{ 10}$,
A.V.\thinspace Dolgopolov$^{  5}$,
S.V.\thinspace Donskov$^{  5}$,
B.C.\thinspace Earl$^{  3}$,
D.\thinspace Evans$^{  3}$,
B.R.\thinspace French$^{  4}$,
T.\thinspace Hino$^{ 11}$,
S.\thinspace Inaba$^{   8}$,
A.V.\thinspace Inyakin$^{  5}$,
T.\thinspace Ishida$^{   8}$,
A.\thinspace Jacholkowski$^{   4}$,
T.\thinspace Jacobsen$^{  10}$,
G.T\thinspace Jones$^{  3}$,
G.V.\thinspace Khaustov$^{  5}$,
T.\thinspace Kinashi$^{  12}$,
J.B.\thinspace Kinson$^{   3}$,
A.\thinspace Kirk$^{   3}$,
W.\thinspace Klempt$^{  4}$,
V.\thinspace Kolosov$^{  5}$,
A.A.\thinspace Kondashov$^{  5}$,
A.A.\thinspace Lednev$^{  5}$,
V.\thinspace Lenti$^{  4}$,
S.\thinspace Maljukov$^{   7}$,
P.\thinspace Martinengo$^{   4}$,
I.\thinspace Minashvili$^{   7}$,
T.\thinspace Nakagawa$^{  11}$,
K.L.\thinspace Norman$^{   3}$,
J.P.\thinspace Peigneux$^{  1}$,
S.A.\thinspace Polovnikov$^{  5}$,
V.A.\thinspace Polyakov$^{  5}$,
V.\thinspace Romanovsky$^{   7}$,
H.\thinspace Rotscheidt$^{   4}$,
V.\thinspace Rumyantsev$^{   7}$,
N.\thinspace Russakovich$^{   7}$,
V.D.\thinspace Samoylenko$^{  5}$,
A.\thinspace Semenov$^{   7}$,
M.\thinspace Sen\'{e}$^{   4}$,
R.\thinspace Sen\'{e}$^{   4}$,
P.M.\thinspace Shagin$^{  5}$,
H.\thinspace Shimizu$^{ 12}$,
A.V.\thinspace Singovsky$^{ 1,5}$,
A.\thinspace Sobol$^{   5}$,
A.\thinspace Solovjev$^{   7}$,
M.\thinspace Stassinaki$^{   2}$,
J.P.\thinspace Stroot$^{  6}$,
V.P.\thinspace Sugonyaev$^{  5}$,
K.\thinspace Takamatsu$^{ 9}$,
G.\thinspace Tchlatchidze$^{   7}$,
T.\thinspace Tsuru$^{   8}$,
M.\thinspace Venables$^{  3}$,
O.\thinspace Villalobos Baillie$^{   3}$,
M.F.\thinspace Votruba$^{   3}$,
Y.\thinspace Yasu$^{   8}$.
}\end{center}

\begin{center}{\bf {{\bf Abstract}}}\end{center}

{
A partial wave analysis of the centrally produced \kk and \ksks
channels has been performed in $pp$ collisions using
an incident beam momentum of 450~GeV/c.
An unambiguous physical solution has been found
in each channel.
The striking feature is the
observation of peaks in the S-wave corresponding to
the $f_0(1500)$ and $f_J(1710)$ with J~=~0.
The D-wave shows evidence for the
$f_2(1270)/a_2(1320)$,
the $f_2^\prime(1525)$ and the $f_2(2150)$ but
there is no evidence for a statistically significant contribution
in the D-wave in the 1.7~GeV mass region.
}
\bigskip
\bigskip
\bigskip
\bigskip\begin{center}{{Submitted to Physics Letters}}
\end{center}
\bigskip
\bigskip
\begin{tabbing}
aba \=   \kill
$^1$ \> \small
LAPP-IN2P3, Annecy, France. \\
$^2$ \> \small
Athens University, Physics Department, Athens, Greece. \\
$^3$ \> \small
School of Physics and Astronomy, University of Birmingham, Birmingham, U.K. \\
$^4$ \> \small
CERN - European Organization for Nuclear Research, Geneva, Switzerland. \\
$^5$ \> \small
IHEP, Protvino, Russia. \\
$^6$ \> \small
IISN, Belgium. \\
$^7$ \> \small
JINR, Dubna, Russia. \\
$^8$ \> \small
High Energy Accelerator Research Organization (KEK), Tsukuba, Ibaraki 305,
Japan. \\
$^{9}$ \> \small
Faculty of Engineering, Miyazaki University, Miyazaki, Japan. \\
$^{10}$ \> \small
Oslo University, Oslo, Norway. \\
$^{11}$ \> \small
Faculty of Science, Tohoku University, Aoba-ku, Sendai 980, Japan. \\
$^{12}$ \> \small
Faculty of Science, Yamagata University, Yamagata 990, Japan. \\
\end{tabbing}
\end{titlepage}
\setcounter{page}{2}
\bigskip
\par
One of the fundamental predictions of QCD is the existence
of glueballs.
Current theoretical predictions based on lattice gauge calculations
indicate that the lowest lying scalar glueball should be in the
mass range 1500-1700~MeV~\cite{re:lgt}.
The $f_0(1500)$ and the
$f_J(1710)$ display clear glueball characteristics
in that they are both produced in glue-rich
production mechanisms and are either not seen, or are
heavily suppressed in normal hadronic interactions.
In addition, the $f_J(1710)$ is observed to decay dominantly to
$K \overline K$ and yet it is not produced in $K^-p$
interactions~\cite{re:LASS}.
However, the spin of the $f_J(1710)$ is still uncertain with J~=~0 or 2
being possible.
\par
In 1987 the MARK III collaboration published the results
of a spin analysis of the $f_J(1710)$ observed in
radiative $J/\psi$ decays~\cite{re:MARK3}. The analysis assumed that
the $f_J(1710)$ region was pure spin zero or pure spin two and
ignored interference effects.
The results from the spin analysis showed that spin two was preferred
over spin zero.
\par
In 1989 the WA76 collaboration published results from the analysis of
the centrally produced $K\overline K$ system~\cite{oldkk}.
In an attempt to assess
the spin of the $f_J(1710)$ the angular distributions in the
1.5 and 1.7 GeV mass regions were studied.
The angular distributions in the
two mass regions were found to be similar, and since it was
assumed that the signal at 1.5 GeV was due to the $f_2^\prime(1525)$, it
was concluded that the signal at 1.7 GeV was also spin two.
{}From these two observations the $f_J(1710)$ was allocated spin two.
\par
A more sophisticated analysis of the Mark III data~\cite{MARK3b}
including both spin zero and two amplitudes
and the possibility of interference between them
showed that the 1.7~GeV mass region was dominated by spin zero.
However, these analyses have only ever been published in conference
proceedings and have never been followed up with publications
in refereed journals.
\par
More recently, the E690 experiment at Fermilab has
published the results of a partial wave analysis of the
centrally produced \ksks system~\cite{E690KK}. The mass spectrum looks very
similar to that of the WA76 experiment; however, the peak at
1.5 GeV is found to have spin zero and not spin two as had been assumed
in the WA76 analysis. Unfortunately, above 1.58 GeV,
due to an ambiguity in the solutions, a unique determination of the
spin of the $f_J(1710)$ could not be made by E690.
\par
It is essential to know the spin of the $f_J(1710)$ because
measurements of its production rate in $J/\psi$ decays indicates that
if it has spin two then it would be consistent with being a normal
$q \overline q$ meson whereas if it has spin zero it would
be consistent with having a significant glueball component~\cite{re:cfl}.
\par
This paper presents a study of the
\kk and \ksks final states formed in the reaction
\begin{equation}
pp \rightarrow p_{f} (K \overline K) p_{s}
\end{equation}
at 450 GeV/c.
The subscripts $f$ and $s$ indicate the
fastest and slowest particles in the laboratory respectively.
The WA102 experiment
has been performed using the CERN Omega Spectrometer,
the layout of which is
described in ref.~\cite{WADPT}.
\par
The reaction
\begin{equation}
pp \rightarrow p_{f} (K^+ K^-) p_{s}
\label{eq:b}
\end{equation}
has been isolated
from the sample of events having four
outgoing
charged tracks,
by first imposing the following cuts on the components of the
missing momentum:
$|$missing~$P_{x}| <  14.0$ GeV/c,
$|$missing~$P_{y}| <  0.16$ GeV/c and
$|$missing~$P_{z}| <  0.08$ GeV/c,
where the x axis is along the beam
direction.
A correlation between
pulse-height and momentum
obtained from a system of
scintillation counters was used to ensure that the slow
particle was a proton.
\par
In order to select the \kk system, information
from the {\v C}erenkov counter was used.
One centrally produced charged
particle was required to be identified as an ambiguous $K$/$p$
by the {\v C}erenkov counter
and the other particle was required to be consistent with being a kaon.
The largest contamination to the real \kk final state comes from the
reaction $pp \rightarrow \Delta^{++}_f p_s \pi^-$
where the high momentum $\pi^+$ from the decay of the $\Delta^{++}$
is misidentified as a $K$ by the {\v C}erenkov system.
In order to reject this contamination the positive particle
was assigned the $\pi$ mass and
the $\Delta^{++}(1232)$ signal has been removed
by requiring $M(p_f\pi^+)$~$>$~1.5~GeV.
\par
The method of Ehrlich et al.~\cite{EHRLICH},
has been used to compute the mass
squared of the two centrally produced particles assuming them to have
equal mass.
The resulting distribution is shown in
fig.~\ref{fi:1}a) where peaks can be seen at
the $\pi$, $K$ and $p$ masses squared.
A cut on the Ehrlich mass squared of
$0.14 \leq M^2_X \leq 0.55$~$GeV^2$
has been used to select a sample of 30~868 \kk events.
\par
The peak at the $\pi$ mass squared results from
$\pi$s being misidentified as $K$s by an inefficiency of the
{\v C}erenkov counter.
To determine the contamination from these \pipi events inside the
Ehrlich mass cut, real \pipi events have been passed through a simulation
of the apparatus in which the efficiency of the
{\v C}erenkov counter (94~$\pm$~1~$\%$) has been taken into account.
The resulting distribution is shown as the
shaded histogram in
fig.~\ref{fi:1}a).
The amount of contamination is small and the
effect this contamination has on the angular distribution
has been found to be negligible~\cite{bethesis}.
\par
The centrally produced \kk effective mass distribution is shown
in fig.~\ref{fi:1}b).
The main features of the spectrum are evidence for a sharp threshold
enhancement and
peaks in the 1.5 and 1.7 GeV regions.
\par
A Partial Wave Analysis (PWA) of the centrally produced \kk system has been
performed assuming the \kk system is produced by the
collision of two particles (referred to as exchanged particles) emitted
by the scattered protons.
The z axis
is defined by the momentum vector of the
exchanged particle with the greatest four-momentum transferred
in the \kk centre of mass.
The y axis is defined
by the cross product of the two exchanged particles in the $pp$ centre of mass.
The two variables needed to specify the decay process were taken as the polar
and azimuthal angles ($\theta$, $\phi$) of the $K^-$ in the \kk
centre of mass relative to the coordinate system described above.
\par
The acceptance corrected moments, defined by
\begin{equation}
I(\Omega) =  \sum _L t_{L0} Y^0 _L(\Omega) + 2 \sum _{L,M >0}
t_{LM}Re\{Y^M_L(\Omega)\}
\end{equation}
have been rescaled to the total number of observed events and
are shown
in fig.~\ref{fi:2}.
As can be seen the moments
with $M$~$>$~2 and $L$~$>$~4 are small (i.e. $t_{43}$, $t_{44}$,
$t_{50}$ and $t_{60}$)
and hence only
S, P, and D waves with $m$~$\leq$~1 have been included in the PWA.
The $t_{00}$ moment represents the total acceptance corrected mass spectrum.
As can be seen, by comparing this spectrum with the raw mass spectrum shown in
fig.~\ref{fi:1}b),
the effect of the acceptance is to increase the proportion of events
with masses less than 1.3~GeV. This is mainly due to the
effects of the momentum threshold of the
{\v C}erenkov counter.
\par
The amplitudes used for the PWA are defined in the reflectivity
basis~\cite{reflectivity}.
In this basis the angular distribution is given by a sum of two non-interfering
terms corresponding to negative and positive values of reflectivity.
The waves used were of the form $J^\varepsilon _m$ with $J$~$=$~S, P and D,
$m$~$=$~$0,1$ and reflectivity $\varepsilon$~=~$\pm 1$.
The expressions relating the moments
($t_{LM}$) and the waves ($J^\varepsilon _m$) are given in table~\ref{ta:a}.
Since the overall phase for each reflectivity is indeterminate,
one wave in each reflectivity can be set to be real ($S_0^-$ and $P_1^+$
for example) and hence two phases can be set to zero ($\phi_{S_0^-}$ and
$\phi_{P_1^+}$ have been chosen).
This results in 12 parameters to be determined from the fit to the
angular distributions.
\par
The PWA has been performed independently in 40~MeV intervals of the \kk
mass spectrum. In each mass bin an event-by-event maximum likelihood
method has been used. The function
\begin{equation}
F=-\sum_{i=1}^Nln\{I(\Omega)\} + \sum_{L,M}t_{LM}\epsilon_{LM}
\end{equation}
has been minimised, where N is the number of events in a given mass bin,
$\epsilon_{LM}$ are the efficiency corrections calculated in the
centre of the bin
and $t_{LM}$ are the moments of the angular distribution.
The moments calculated from the partial amplitudes
are shown superimposed on the experimental moments
in fig~\ref{fi:2}. As can be seen the results of the fit
reproduce well the experimental moments.
\par
The equations that express the moments via the partial wave
amplitudes form a non-linear system that leads to inherent
ambiguities. For a system with S, P and D waves there are eight solutions
for each mass bin.
In each mass bin
one of these solutions
is found from the fit to the experimental angular distributions;
the other seven can then be calculated by the method described in
ref.~\cite{reflectivity}.
In order to link the solutions in adjacent mass bins, the real and
imaginary parts of the Barrelet function roots are required to be
step-wise
continuous and have finite derivatives as a function of mass~\cite{link}.
By definition all the solutions give identical moments and identical
values of the likelihood.
The only way to differentiate between the solutions, if different, is to
apply some external physical test, such as requiring
that at threshold the S-wave is the dominant wave.
\par
The four complex roots, Z$_i$, after the linking procedure are shown in
fig.~\ref{fi:1}c) and d).
As can be seen
the imaginary parts give little help in the linking procedure. However,
the real parts are
well separated in most places
and hence it is possible to identify unambiguously
all the PWA solutions in the whole mass range.
In the 1.8 GeV mass region two roots do become close together, but
if these two roots are swapped it results in events from the
S-wave being transferred almost entirely into the P-wave and produces
mass spectra that look unphysical.
In addition,
the zeros do not cross the real axis
and hence there is no problem with bifurcation of the solutions.
Near threshold the P-wave is the dominant contribution for five out
of eight solutions, another is dominated by D-wave and
another has the same amount of S-wave and P-wave. These
seven solutions~\cite{bethesis}
have been ruled out because the \kk cross section near threshold has been
assumed to be dominated by S-wave.
The remaining solution is shown in fig.~\ref{fi:3}.
\par
The S-wave shows a threshold enhancement; the peaks at 1.5 GeV and
1.7~GeV are interpreted as being due to the
$f_0(1500)$ and $f_J(1710)$ with J~=~0.
The D-wave shows peaks in the 1.3 and 1.5~GeV regions,
presumably due to the $f_2(1270)/a_2(1320)$ and $f_2^\prime(1525)$ and
a wide structure above 2 GeV. There is no evidence for
any significant structure in the D-wave in the region of the
$f_J(1710)$. In addition, there are no statistically significant
structures in any of the other waves.
\par
Extensive Monte Carlo simulations have been performed to check the
validity of this result.
They show that the feed through from the S-wave to the D-wave is
approximately 5~\% from threshold to 1.3 GeV decreasing to less
than 1 \% for masses greater than 1.5 GeV.
The feed through from D-wave to S-wave is found to be
negligible over the entire mass range and hence gives confidence
in the above observation that the $f_J(1710)$ has J~=~0.
\par
A fit has been performed to the $S_0^-$ wave using
three interfering Breit-Wigners to describe the $f_0(980)$, $f_0(1500)$
and $f_J(1710)$ and a background
of the form
$a(m-m_{th})^{b}exp(-cm-dm^{2})$, where
$m$ is the
\kk
mass,
$m_{th}$ is the
\kk
threshold mass and
a, b, c, d are fit parameters.
The Breit-Wigners have been convoluted with a Gaussian to
account for the experimental mass resolution
($\sigma$~=~6 MeV at threshold rising to 19~MeV at 2~GeV).
The resulting fit is shown in fig.~\ref{fi:4}a) and gives
\begin{tabbing}
aaaa\=adfsfsf99ba \=Mas \= == \=1224 \=pm \=1200 \=MeVswfw, \=gaa \=  == \=1224
\=pm \=1200  \=MeV   \kill
\>$f_0(980)$ \>M \>=\>985\>$\pm$\>10\>MeV,\>$\Gamma$\>=\>65\>$\pm$\>20\>MeV \\
\>$f_0(1500)$ \>M \>=\>1497\>$\pm$\>10\>MeV,\>$\Gamma$\>=\>104\>$\pm$\>25\>MeV
\\
\>$f_0(1710)$ \>M \>=\>1730\>$\pm$\>15\>MeV,\>$\Gamma$\>=\>100\>$\pm$\>25\>MeV
\end{tabbing}
parameters which are consistent with the PDG~\cite{PDG98} values for these
resonances.
\par
A fit has been performed to the $D_0^-$ wave above 1.2~GeV using
three incoherent relativistic spin 2 Breit-Wigners
to describe the $f_2(1270)/a_2(1320)$,
$f_2^\prime(1525)$ and the peak at 2.2 GeV
a background
of the form
$a(m-m_{th})^{b}exp(-cm-dm^{2})$, where
$m$ is the
\kk
mass,
$m_{th}$ is the
\kk
threshold mass and
a, b, c, d are fit parameters.
The resulting fit is shown in fig.~\ref{fi:4}b) and gives
\begin{tabbing}
aaaa\=adfsfsf9sefwefef9ba \=Mas \= == \=1224 \=pm \=1200 \=MeVswfw, \=gaa \=
== \=1224 \=pm \=1200  \=MeV   \kill
\>$f_2(1270)/a_2(1320)$ \>M
\>=\>1305\>$\pm$\>20\>MeV,\>$\Gamma$\>=\>132\>$\pm$\>25\>MeV \\
\>$f_2^\prime(1525)$ \>M
\>=\>1515\>$\pm$\>15\>MeV,\>$\Gamma$\>=\>70\>$\pm$\>25\>MeV \\
\>$f_2(2150)$ \>M \>=\>2130\>$\pm$\>35\>MeV,\>$\Gamma$\>=\>270\>$\pm$\>50\>MeV.
\end{tabbing}
The parameters for the peak at 1.3 GeV fall between the PDG~\cite{PDG98}
values for the
$f_2(1270)$ and $a_2(1320)$.
The values for the $f_2^\prime(1525)$ are compatible with the PDG values
and those for the structure at 2.15 GeV are compatible with both
the $f_2(2150)$ and with the
parameters for the structure seen in radiative $J/\psi$ decays
to the same channel~\cite{JPSI21}.
\par
A study has also been made of the centrally produced \ksks channel.
This channel has lower statistics than the \kk channel but has
the advantage that only even spins can contribute, which also means that
there are only two ambiguous solutions to the PWA.
The reaction
\begin{equation}
pp \rightarrow p_{f} (K^0_S K^0_S) p_{s}
\label{eq:e}
\end{equation}
has been isolated
from the sample of events having two
outgoing
charged tracks plus two $V^0$s,
by first imposing the following cuts on the components of the
missing momentum:
$|$missing~$P_{x}| <  20.0$ GeV/c,
$|$missing~$P_{y}| <  0.16$ GeV/c and
$|$missing~$P_{z}| <  0.12$ GeV/c.
\par
The quantity $\Delta$, defined as
$ \Delta = MM^{2}(p_{f}p_{s}) - M^{2}(K^0_S K^0_S)$,
where $MM^{2}(p_{f}p_{s})$ is the missing mass squared of the two
outgoing protons,
was then calculated for each event and
a cut of $|\Delta|$ $\leq$ 3.0 (GeV)$^{2}$ was used to select the
\ksks channel.
Requiring one $V^0$ to be compatible with being a $K^0_S$
(0.475~$<$~$M(\pi^+\pi^-)$~$<$~0.520~GeV)
the effective mass of the other $V^0$
is shown in fig~\ref{fi:5}a) where
a clear $K^0_S$ signal can be seen
over little background.
The resulting \ksks effective mass spectrum is shown in fig.~\ref{fi:5}b)
and consists of 2712 events.
\par
A Partial Wave Analysis (PWA) of the centrally produced \ksks system has
then been
performed assuming the same analysis frame as for the \kk system.
Since only even spins can contribute only
S and D waves with $m$~$\leq$~1 have been included in the PWA.
The expressions relating the moments
and the waves are given in table~\ref{ta:b}.
The PWA has been performed independently in 80~MeV intervals of the \ksks
mass spectrum.
\par
The two complex roots, Z$_i$, after the linking procedure, are shown in
fig.~\ref{fi:5}c) and d).
As in the case of the \kk channel
the real parts are
well separated
and hence it is possible to identify unambiguously
the two PWA solutions in the whole mass range.
In addition,
the zeros do not cross the real axis
and hence there is no problem with bifurcation of the solutions.
For one solution the mass spectrum is evenly distributed through the
three D-waves and the S-wave is small everywhere, this solution has
been ruled out.
The remaining solution is shown in fig.~\ref{fi:5}e)-h).
As in the case of the \kk final state,
it can be seen that
the S-wave shows a threshold enhancement; the peaks at 1.5 GeV and
1.7~GeV are interpreted as being due to the
$f_0(1500)$ and $f_J(1710)$ with J~=~0.
Superimposed on the S-wave is the result of a fit using the
parametrisation used to fit the \kk S-wave.
As can be seen this parametrisation describes the \ksks S-wave.
\par
In conclusion, a partial wave analysis of the centrally
produced \kk and \ksks systems has been performed.
An unambiguous physical solution has been found
in each channel.
The striking feature is the
observation of peaks in the S-wave corresponding to
the $f_0(1500)$ and $f_J(1710)$ with J~=~0.
The D-wave shows evidence for the
$f_2(1270)/a_2(1320)$,
the $f_2^\prime(1525)$ and the $f_2(2150)$
but there is no evidence for a statistically significant contribution
in the D-wave in the 1.7~GeV mass region.
\begin{center}
{\bf Acknowledgements}
\end{center}
\par
This work is supported, in part, by grants from
the British Particle Physics and Astronomy Research Council,
the British Royal Society,
the Ministry of Education, Science, Sports and Culture of Japan
(grants no. 04044159 and 07044098), the Programme International
de Cooperation Scientifique (grant no. 576)
and
the Russian Foundation for Basic Research
(grants 96-15-96633 and 98-02-22032).

\newpage
\newpage
\begin{table}[h]
\caption{The moments of the angular distribution expressed in terms of the
partial waves for the \kk system.}
\label{ta:a}
\vspace{1in}
\begin{center}
\begin{tabular}{|ccl|} \hline
  & & \\
 $\sqrt{4 \pi}t_{00}$ & = &
$|S_0^-|^2+|P_0^-|^2+|P^-_1|^2+|P^+_1|^2+|D_0^-|^2+|D^-_1|^2+|D^+_1|^2$\\
  & & \\
 $\sqrt{4 \pi}t_{10}$ & = &
$2|S_0^-||P_0^-|cos(\phi_{S_0^-}-\phi_{P_0^-})+
\frac{4}{\sqrt{5}}|P_0^-||D_0^-|cos(\phi_{P_0^-}-\phi_{D_0^-})$\\
&&$
+\frac{2\sqrt{3}}{\sqrt{5}}\{|P_1^-||D_1^-|cos(\phi_{P_1^-}-
\phi_{D_1^-})+|P_1^+||D_1^+|cos(\phi_{P_1^+}-\phi_{D_1^+})\}$\\
  & & \\
 $\sqrt{4 \pi}t_{11}$ & = &
$\sqrt{2}|S_0^-||P_1^-|cos(\phi_{S_0^-}-\phi_{P_1^-})-
\frac{\sqrt{2}}{\sqrt{5}}|P_1^-||D_0^-|cos(\phi_{P_1^-}-\phi_{D_0^-})$\\
&&$ +\frac{\sqrt{6}}{\sqrt{5}}|P_0^-||D_1^-|cos(\phi_{P_0^-}-\phi_{D_1^-})$\\
  & & \\
 $\sqrt{4 \pi}t_{20}$ & = &
$\frac{2}{\sqrt{5}}|P_0^-|^2-\frac{1}{\sqrt{5}}(|P_1^-|^2+|P_1^+|^2)
+\frac{\sqrt{5}}{7}(2|D_0^-|^2 + |D_1^-|^2 + |D_1^+|^2)$\\
&&$ +2|S_0^-||D_0^-|cos(\phi_{S_0^-}-\phi_{D_0^-})$\\
  & & \\
 $\sqrt{4 \pi}t_{21}$ & = &
$\frac{\sqrt{6}}{\sqrt{5}}|P_1^-||P_0^-|cos(\phi_{P_1^-}-
\phi_{P_0^-})+\frac{\sqrt{10}}{7}|D_1^-||D_0^-|cos(\phi_{D_1^-}-
\phi_{D_0^-})$\\
&&$ +\sqrt{2}|S_0^-||D_1^-|cos(\phi_{S_0^-}-\phi_{D_1^-})$\\
  & & \\
 $\sqrt{4 \pi}t_{22}$ & = &
$\frac{\sqrt{3}}{\sqrt{10}}(|P_1^-|^2-|P_1^+|^2)+
\frac{\sqrt{15}}{7\sqrt{2}}(|D_1^-|^2-|D_1^+|^2)$\\
  & & \\
 $\sqrt{4 \pi}t_{30}$ & = &$
-\frac{6}{\sqrt{35}}\{|P_1^-||D_1^-|cos(\phi_{P_1^-}-
\phi_{D_1^-})+|P_1^+||D_1^+|cos(\phi_{P_1^+}-\phi_{D_1^+})\}$\\
 &&
$+\frac{6\sqrt{3}}{\sqrt{35}}|P_0^-||D_0^-|cos(\phi_{P_0^-}-\phi_{D_0^-})$\\
  & & \\
 $\sqrt{4 \pi}t_{31}$ & = &
$\frac{6}{\sqrt{35}}|P_1^-||D_0^-|cos(\phi_{P_1^-}-
\phi_{D_0^-})+\frac{4\sqrt{3}}{\sqrt{35}}|P_0^-
|D_1^-|cos(\phi_{P_0^-}-\phi_{D_1^-})$\\
  & & \\
 $\sqrt{4 \pi}t_{32}$ & = &
$\frac{\sqrt{6}}{\sqrt{7}}\{|P_1^-||D_1^-|cos(\phi_{P_1^-}-
\phi_{D_1^-})-|P_1^+||D_1^+|cos(\phi_{P_1^+}-\phi_{D_1^+})\}$\\
  & & \\
 $\sqrt{4 \pi}t_{40}$ & = &
$\frac{6}{7}|D_0^-|^2-\frac{4}{7}(|D_1^-|^2+|D_1^+|^2)$\\
  & & \\
 $\sqrt{4 \pi}t_{41}$ & = &
$\frac{2\sqrt{15}}{7}|D_0^-||D_1^-|cos(\phi_{D_0^-}-\phi_{D_1^-})$\\
  & & \\
 $\sqrt{4 \pi}t_{42}$ & = & $\frac{\sqrt{10}}{7}(|D_1^-|^2-|D_1^+|^2)$\\
  & & \\ \hline
\end{tabular}
\end{center}
\end{table}
\newpage
\begin{table}[h]
\caption{The moments of the angular distribution expressed in terms of the
partial waves for the \ksks system.}
\label{ta:b}
\vspace{1in}
\begin{center}
\begin{tabular}{|ccl|} \hline
  & & \\
 $\sqrt{4 \pi}t_{00}$ & = & $|S_0^-|^2+|D_0^-|^2+|D^-_1|^2+|D^+_1|^2$\\
  & & \\
 $\sqrt{4 \pi}t_{20}$ & = & $\frac{\sqrt{5}}{7}(2|D_0^-|^2 + |D_1^-|^2 +
|D_1^+|^2)$\\
&&$ +2|S_0^-||D_0^-|cos(\phi_{S_0^-}-\phi_{D_0^-})$\\
  & & \\
 $\sqrt{4 \pi}t_{21}$ & = &
$\frac{\sqrt{10}}{7}|D_1^-||D_0^-|cos(\phi_{D_1^-}-\phi_{D_0^-})$\\
&&$ +\sqrt{2}|S_0^-||D_1^-|cos(\phi_{S_0^-}-\phi_{D_1^-})$\\
  & & \\
 $\sqrt{4 \pi}t_{22}$ & = &
$\frac{\sqrt{15}}{7\sqrt{2}}(|D_1^-|^2-|D_1^+|^2)$\\
  & & \\
 $\sqrt{4 \pi}t_{40}$ & = &
$\frac{6}{7}|D_0^-|^2-\frac{4}{7}(|D_1^-|^2+|D_1^+|^2)$\\
  & & \\
 $\sqrt{4 \pi}t_{41}$ & = &
$\frac{2\sqrt{15}}{7}|D_0^-||D_1^-|cos(\phi_{D_0^-}-\phi_{D_1^-})$\\
  & & \\
 $\sqrt{4 \pi}t_{42}$ & = & $\frac{\sqrt{10}}{7}(|D_1^-|^2-|D_1^+|^2)$\\
  & & \\ \hline
\end{tabular}
\end{center}
\end{table}
\clearpage
{ \large \bf Figures \rm}
\begin{figure}[h]
\caption{a) The Ehrlich mass squared distribution and
b) the \kk mass spectrum.
The c) Real and d) Imaginary parts of the roots (see text)
as a function of mass obtained from the PWA of the \kk system.
}
\label{fi:1}
\end{figure}
\begin{figure}[h]
\caption{ The $\protect\sqrt{4\pi}t_{LM}$ moments from the data.
are the resulting
moments calculated from the PWA of the
\kk final state.
}
\label{fi:2}
\end{figure}
\begin{figure}[h]
\caption{The physical solution from the PWA of the \kk final state.
}
\label{fi:3}
\end{figure}
\begin{figure}[h]
\caption{The a) $S_0^-$ and b) $D_0^-$ waves with fits described in the text.
}
\label{fi:4}
\end{figure}
\begin{figure}[h]
\caption{
The $K^0_SK^0_S$ final state.
a) the $M(\pi^+\pi^-$) mass spectrum when the other $V^0$ is compatible
with being a $K^0_S$.
b) The \ksks mass spectrum.
The c) Real and d) Imaginary parts of the roots (see text)
as a function of mass obtained from the PWA of the \ksks system.
e)-h) The physical solution from the PWA of the \ksks system.
}
\label{fi:5}
\end{figure}
\newpage
\begin{center}
\epsfig{figure=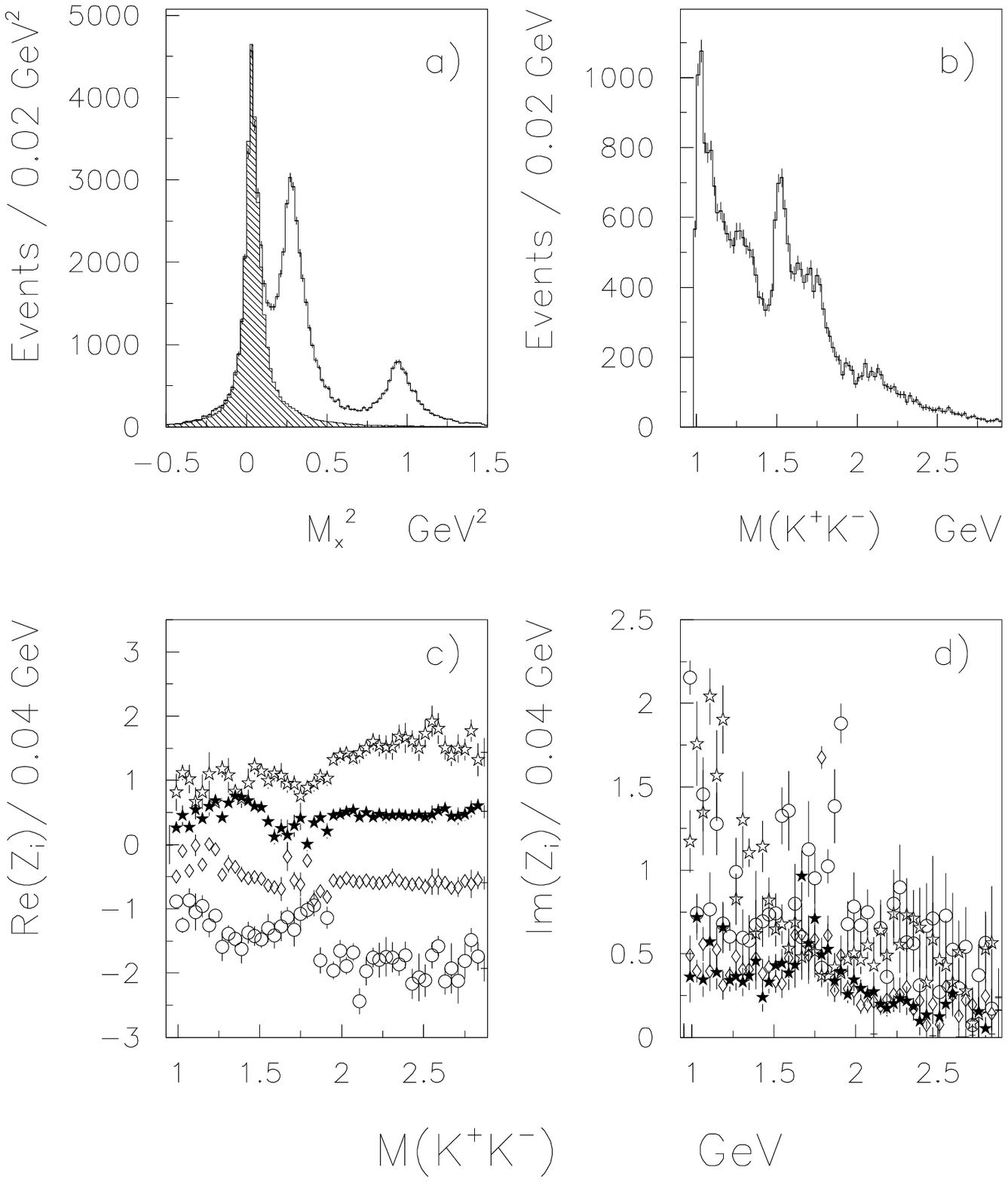,height=22cm,width=17cm}
\end{center}
\begin{center} {Figure 1} \end{center}
\newpage
\begin{center}
\epsfig{figure=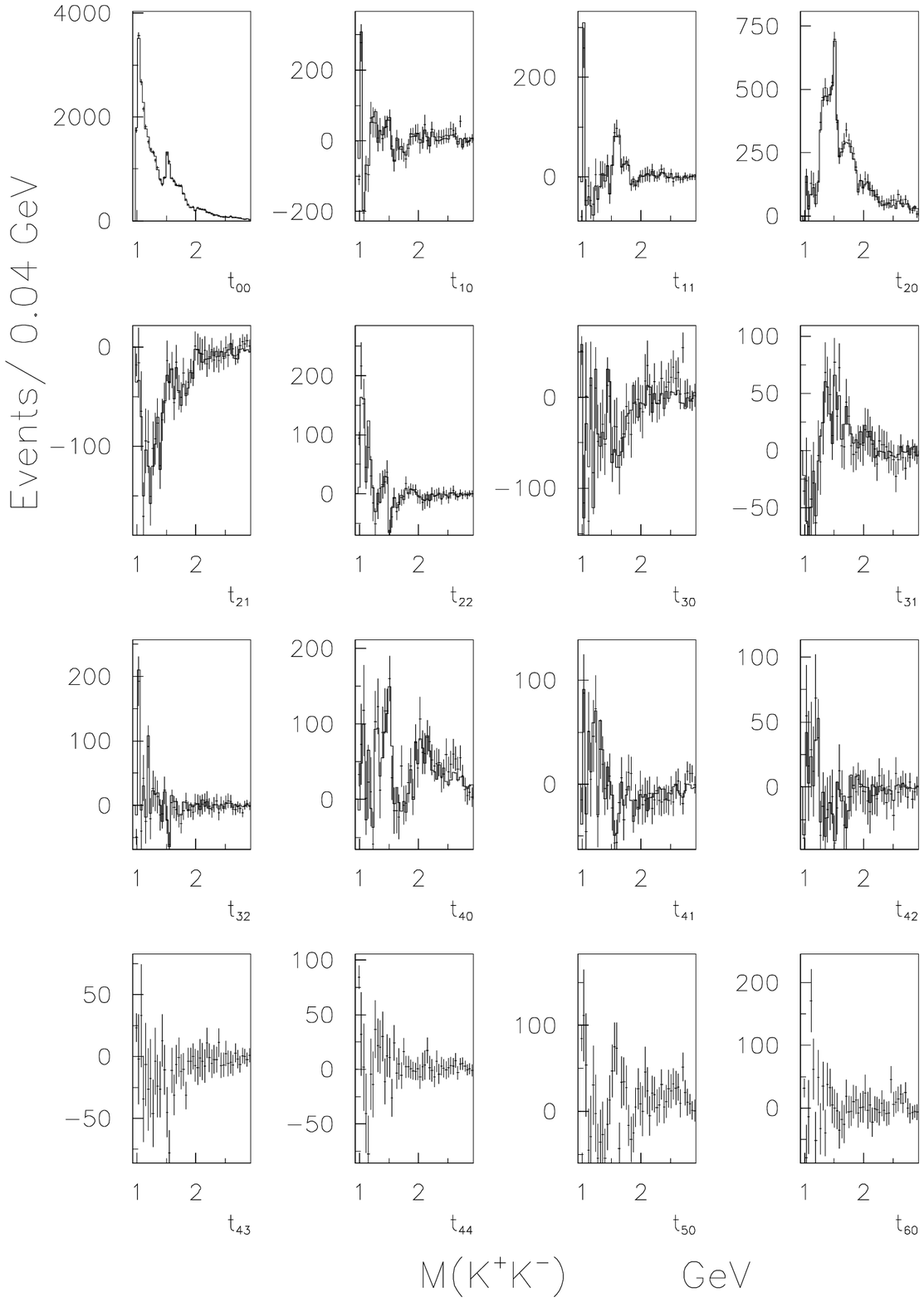,height=22cm,width=17cm,bbllx=0pt,
bblly=0pt,bburx=450pt,bbury=657pt}
\end{center}
\begin{center} {Figure 2} \end{center}
\newpage
\begin{center}
\epsfig{figure=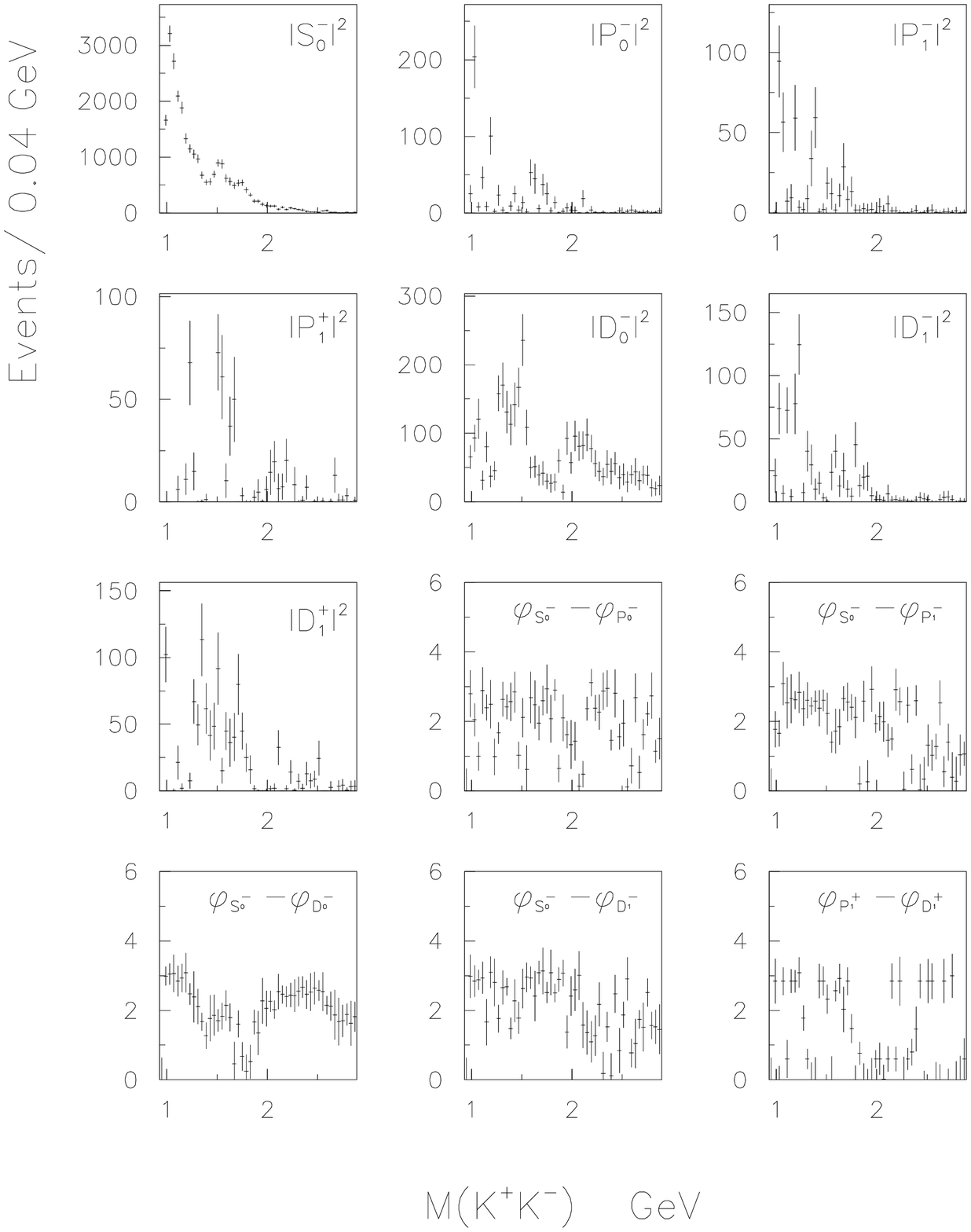,height=22cm,width=17cm,bbllx=0pt,
bblly=0pt,bburx=550pt,bbury=750pt}
\end{center}
\begin{center} {Figure 3} \end{center}
\newpage
\begin{center}
\epsfig{figure=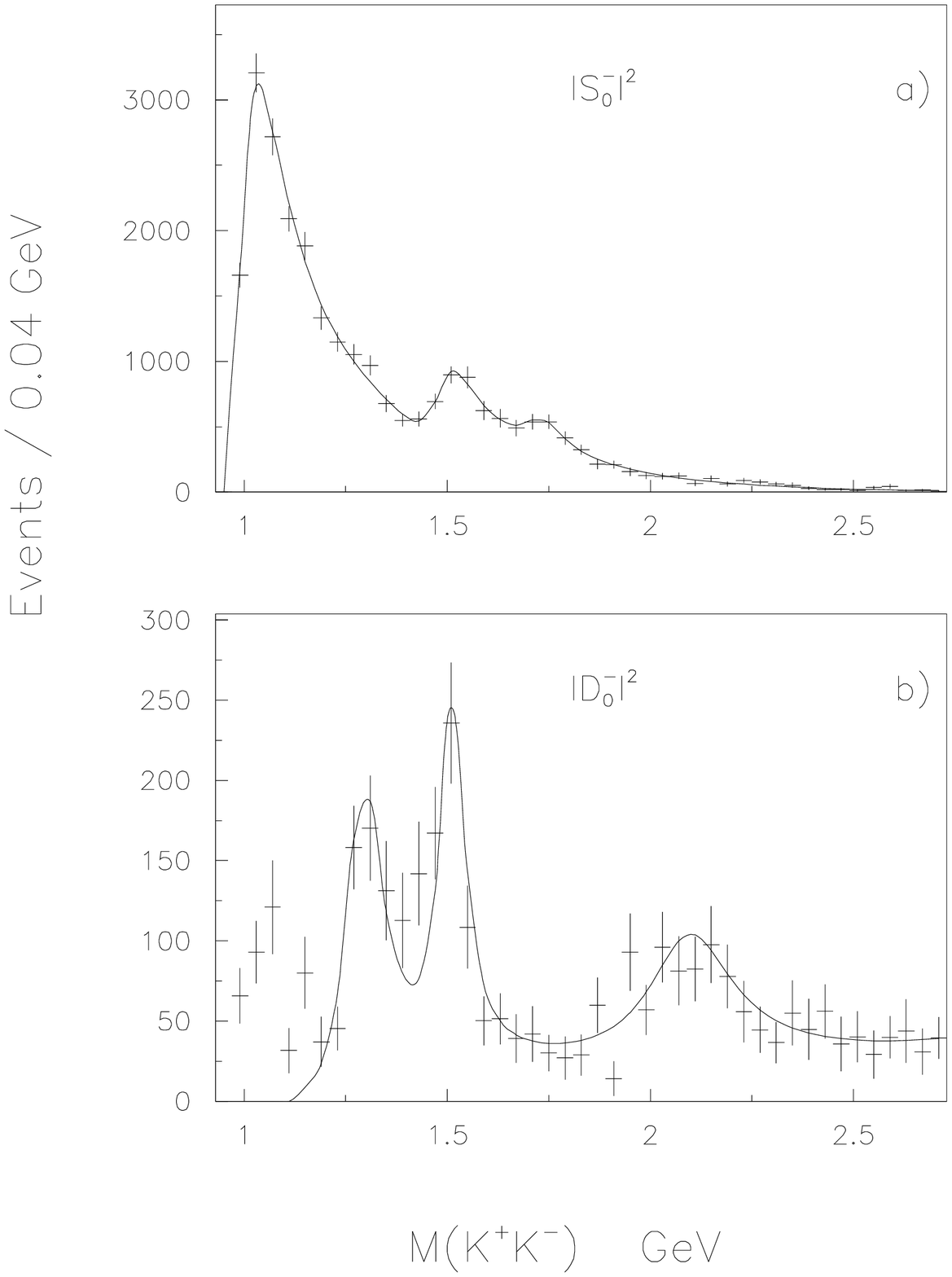,height=22cm,width=17cm,
bbllx=0pt,bblly=0pt,bburx=550pt,bbury=700pt}
\end{center}
\begin{center} {Figure 4} \end{center}
\newpage
\begin{center}
\epsfig{figure=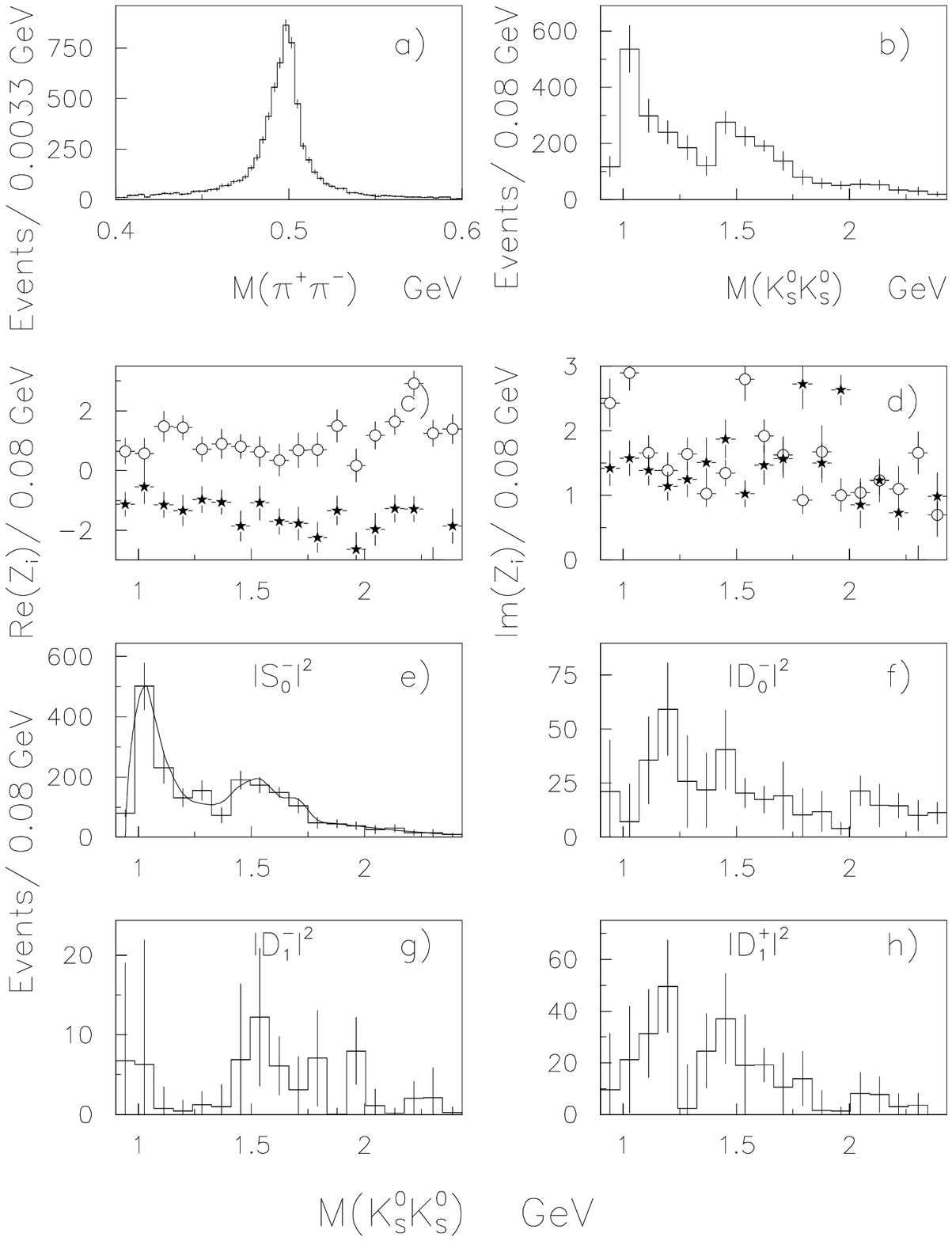,height=22cm,width=17cm,
bbllx=0pt,bblly=0pt,bburx=550pt,bbury=700pt}
\end{center}
\begin{center} {Figure 5} \end{center}

\bigskip
\newpage
\begin{thebibliography}{99}
\bibitem{re:lgt}
G. Bali  {\em et al.,}
Phys. Lett.  {\bf B309}  (1993) 378;

D. Weingarten, hep-lat/9608070;

J. Sexton {\em et al.,} Phys. Rev. Lett. {\bf 75} (1995) 4563;

F.E. Close and M.J. Teper, ``On the lightest Scalar Glueball"
Rutherford Appleton Laboratory report no. RAL-96-040;
Oxford University report no. OUTP-96-35P
\bibitem{re:LASS}
Ph. Gavillet {\em et al.,} Zeit. Phys. {\bf C16} (1982) 119;\\
D. Aston {\em et al.,} Nucl. Phys. {\bf  B301} (1988) 525.
\bibitem{re:MARK3}
R.M. Baltrusaitis {\em et al.,} Phys. Rev.  {\bf D35} (1987) 2077.
\bibitem{oldkk}
T.A. Armstrong {\em et al.,} Phys. Lett.  {\bf B227} (1989) 186.
\bibitem{MARK3b}
L.P. Chen {\em et al.,} Nucl. Phys. Proc. Suppl. {\bf B21} (1991) 80; \\
L.P. Chen {\em et al.,} Proceedings of Hadron 91, Maryland, USA (1991) 26.; \\
W. Dunwoodie, Proceedings of Hadron 97, AIP Conf. Series 432 (1997) 753.
\bibitem{E690KK}
M.A. Reyes {\em et al.,} Phys. Rev. Lett. {\bf 81} (1998) 4079.
\bibitem{re:cfl}
F.E. Close, G.R. Farrar and Z. Li, Phys. Rev. {\bf D55} (1997) 5749.
\bibitem{WADPT}
D. Barberis {\em et al.,} Phys. Lett. {\bf B397 } \rm (1997) 339.
\bibitem{EHRLICH}
R. Ehrlich {\em et al.,} Phys. Rev. Lett. {\bf 20} (1968) 686.
\bibitem{bethesis}
B.C. Earl, Thesis, University of Birmingham, U.K. 1999.
\bibitem{reflectivity}
S.U. Chung, Phys. Rev. {\bf D56 } \rm (1997) 7299.
\bibitem{link}
D. Alde {\em et al.,}, Europ. Phys. Journal. {\bf A3 } \rm (1998) 361.
\bibitem{PDG98}
Particle Data Group, European Physical Journal {\bf C3} (1998) 1.
\bibitem{JPSI21}
R.M. Baltrusaitis {\em et al.,} Phys. Rev. Lett. {\bf 56} (1986) 107; \\
J.E. Augustin {\em et al.,} Phys. Rev. Lett. {\bf 60} (1988) 2238; \\
L. Kopke and N.Wermes, Phys. Rep. {\bf 174} (1989) 67.
\end{thebibliography}
\end{document}